\documentclass[%
 aip,
 amsmath,amssymb,
 reprint,%
]{revtex4-1}

\usepackage{graphicx}
\usepackage{dcolumn}
\usepackage{bm}
\usepackage{color}

\usepackage[utf8]{inputenc}
\usepackage[T1]{fontenc}
\usepackage{mathptmx}

\begin{document}

\preprint{AIP/123-QED}

\title[Chaotic dynamics of graphenes]{Chaotic dynamics of graphene and graphene nanoribbons}

\author{M. Hillebrand}%
\email{hllmal004@myuct.ac.za}
\affiliation{%
Department of Mathematics and Applied Mathematics, University of Cape Town, Rondebosch, 7701, Cape Town, South Africa
}
\author{B. Many Manda}
\email{mnyber004@myuct.ac.za}
\affiliation{%
Department of Mathematics and Applied Mathematics, University of Cape Town, Rondebosch, 7701, Cape Town, South Africa
}

\author{G. Kalosakas}%
\email{georgek@upatras.gr}
\affiliation{
Department of Materials Science, University of Patras, GR-26504 Rio, Greece
}%

\author{E. Gerlach}%
\email{enrico.gerlach@tu-dresden.de}
\affiliation{
Lohrmann Observatory, Technische Universit\"at Dresden, D-01062 Dresden, Germany
}%

\author{Ch. Skokos}
\email{haris.skokos@uct.ac.za}
\affiliation{%
Department of Mathematics and Applied Mathematics, University of Cape Town, Rondebosch, 7701, Cape Town, South Africa
}%

\date{\today}

\begin{abstract}
We study the chaotic dynamics of graphene structures, considering both a periodic, defect free, graphene sheet and graphene nanoribbons (GNRs) of various widths.
By numerically calculating the maximum Lyapunov exponent, we quantify the chaoticity for a spectrum of energies in both systems.
We find that for all cases, the chaotic strength increases with the energy density,
and that the onset of chaos in graphene is slow, becoming evident  after more than $10^4$ natural oscillations of the system.
For the GNRs, we also investigate the impact of the width and chirality (armchair or zigzag edges) on their chaotic behavior.
Our results suggest that due to the free edges the chaoticity of GNRs is stronger than the periodic graphene sheet, and decreases by increasing width, tending asymptotically to the bulk value.
In addition, the chaotic strength of armchair GNRs is higher than a zigzag ribbon of the same width.
Further, we show that the composition of ${}^{12}C$ and ${}^{13}C$ carbon isotopes in graphene has a minor impact on its chaotic strength.
\end{abstract}

\maketitle

\begin{quotation}
In spite of the significance of graphene dynamics in determining a number of structural, conformational, thermal, and vibrational
characteristics of the well studied nanomaterial, which is known to exhibit exceptional properties, a detailed investigation
of its chaotic behavior is missing. Using a realistic, specifically designed, Hamiltonian for the description of the stretching
of covalent bonds and the angle bending in planar graphenes, derived through accurate calculations from first principles
and tested against available experimental data of graphene's mechanical response, we examine the maximum Lyapunov
exponent of two-dimensional bulk graphene as well as of graphene nanoribbons.
In the former case the dependence of this chaos index on the energy of graphene is calculated, revealing a quadratic variation.
In the latter case the dependence of the maximum Lyapunov exponent for both armchair and zigzag nanoribbons on the ribbon
width is presented for various energies and the corresponding results are quantified by a simple analytical function.
\end{quotation}

\section{\label{sec:introduction} Introduction}

Graphene is a defect free single layer of graphite~\cite{NGMJZDGF04}.
It can be considered as the basic building block of most carbon nanomaterials such as graphene nanoribbons (GNR),
carbon nanotubes (CNT) and fullerenes among others.
Experimental studies suggest that graphene carries exceptional physical properties~\cite{RMP09,ZMCLSPR10,SJZDKS11} including superior electron mobility~\cite{NGMJKGDF05,GN07}, thermal conductivity~\cite{BGBCTML08,GCTPNBBM08} and mechanical characteristics~\cite{LWKH08,LYC12,TPPJFGNG09,papagACSNano10,revJPCM15}.
Consequently, it is thought of as a serious candidate for next generation electrodes~\cite{BKB12}, sensors~\cite{sens,YYZYC11}, resonators~\cite{reson}, transistors~\cite{avouris,MCKLPCA12} and super capacitors~\cite{LYNZJ10}.
With its flexibility,  graphene sheets can also be used as an artificial membrane in biomedical research.
Functionalized graphene derivatives open even larger horizons of applications -- for example, applications to drug delivery processes of complex diseases~\cite{ZXZLZ10}.
As such, there is clearly ample motivation for studying various properties of graphene.

Aside from the active experimental research on the diverse applications mentioned above, graphene dynamics can be used
in order to better understand its mechanical and structural properties.
Over the past decades the theoretical investigation of graphene structural properties has resulted in numerous works using primarily molecular dynamics (MD) simulations.
The impact of geometrical factors on graphene's conformation has been investigated by MD~\cite{ZM10}, as well as
the effects of different kinds of defects on the spontaneous formation of various graphene nanostructures, ranging from
GNRs and nanoscrolls ~\cite{nscrolls}, to CNTs~\cite{SSPG14}, nanocages~\cite{ncages} and other exotic nanomaterials~\cite{SKSP15}.

Quantitative analysis of the thermal conductivity, which is an important aspect, is also the subject of extensive study using MD~\cite{GZG09,HRC09,RuoffNatMat12,CK12,ZD12,KSLCBT13,LCYZ13,YLZYW13,YZ13,ZD14,ZHY15}.
For GNRs it has been found that the thermal conductivity increases with their length \cite{GZG09,YZ13} and decreases with
both tensile or compressive uniaxial strain~\cite{GZG09}, while it seems to be insensitive to bending or twisting deformations~\cite{YZ13}.
In addition, MD simulations have shown that rough edges~\cite{SKH10} and vacancies~\cite{LT12} significantly reduce the thermal conductivity of a GNR. MD has been further used to examine other thermal properties of graphene, like its thermal expansion~\cite{IMLB13}.

Phonons in graphene have been also analyzed by MD calculations. In particular it has been shown that the phonon dispersion
curves of graphene can be obtained at any temperature through the dynamical trajectories of the system~\cite{KKGP15}.
Phonon frequency variation with temperature for particular Raman active modes has also been computed \cite{kopidakis,KKGP15}.
Further, the strain dependence of the same phonon of interest ($E_{2g}$ mode) has been dynamically calculated~\cite{AKPKPG15}.
In addition, phonon lifetimes of various optical or acoustic modes were obtained using MD~\cite{phlifetime}.

Concerning the mechanical response of graphene, various elastic parameters, such as the Young modulus, bulk modulus, Poisson ratio, shear modulus, fracture toughness, and critical compressive buckling, have been calculated using MD~\cite{ZMA09,TT10,CQ13,KLGP13,ZMFZPLLGZZAZL14,sgouros2dmat}.
The reported values of the Young modulus are in agreement with the experimental estimates~\cite{LWKH08},
demonstrating the exceptional mechanical properties of graphene.
GNRs' Young modulus (Poisson ratio) increases (decreases) with their size~\cite{ZMA09} regardless of
the chirality (zigzag or armchair). Their Young modulus drops with increased concentration of vacancies~\cite{NP10}.
MD simulations show that GNRs under uniaxial compression exhibit a critical buckling stress which decreases with the
length, while it increases with the width approaching a limiting value at relatively large widths~\cite{SKPG18}.

To the best of our knowledge, there are no studies that have been done regarding the calculation of the maximum Lyapunov exponent (MLE) or other types of chaos indicators~\cite{SGL16} for graphene models, although there is a fundamental interest in understanding the underlying dynamics and the characteristics of chaos in this material.
In this work we assess the stability of a graphene system by computing its MLE.
The MLE has a significant number of applications in complex systems~\cite{S10,PP16}, like for example in detecting phase transitions of matter, and therefore serving as a dynamical order parameter.
For instance in DNA chains, the MLE acts as a dynamical indicator of the phase transition near denaturation~\cite{BD01,HKSS19}.
In practice, for the graphene shell MLE computations could help to detect a threshold above which fractures and deformations appear, as well as to potentially assess the stability of nanomaterials obtained through controllable defect engineering.
In our study we focus on planar graphene sheets and GNRs, modeling the interatomic interactions using a simple graphene-specific two-dimensional Hamiltonian model, which takes into account anharmonic effects~\cite{KLGP13}. The used force fields describe bond stretching
and angle bending interactions, constituting the two-dimensional part of a complete atomistic potential of graphene that has been
derived through accurate calculations from first principles~\cite{PCCP,EPJB}. It has been shown that the obtained in-plane dynamics
accurately describes graphene's mechanical properties~\cite{KLGP13}.
We provide a detailed investigation of the chaoticity of graphene and GNRs through numerical calculations of the system's MLE.
In particular we present the energy dependence of the MLE for periodic graphene, while for GNRs we discuss the width dependence
and the effect of the chirality (armchair or zigzag edges) on the chaotic behavior.

The paper is organized as follows. In section \ref{sec:model} we outline the structure of graphene, the model used, and the computational and numerical tools applied to the problem.
In section \ref{sec:results} we discuss the results of our numerical simulations.
Finally, in section \ref{sec:conclusion} we summarize the main outcomes of our study and discuss the future outlook of our work.

\section{\label{sec:model} Hamiltonian model and computational aspects}

\subsection{\label{subsec:lattice_of_carbon} Graphene lattice}

\begin{figure}
    \centering
    \includegraphics[width=0.48\textwidth]{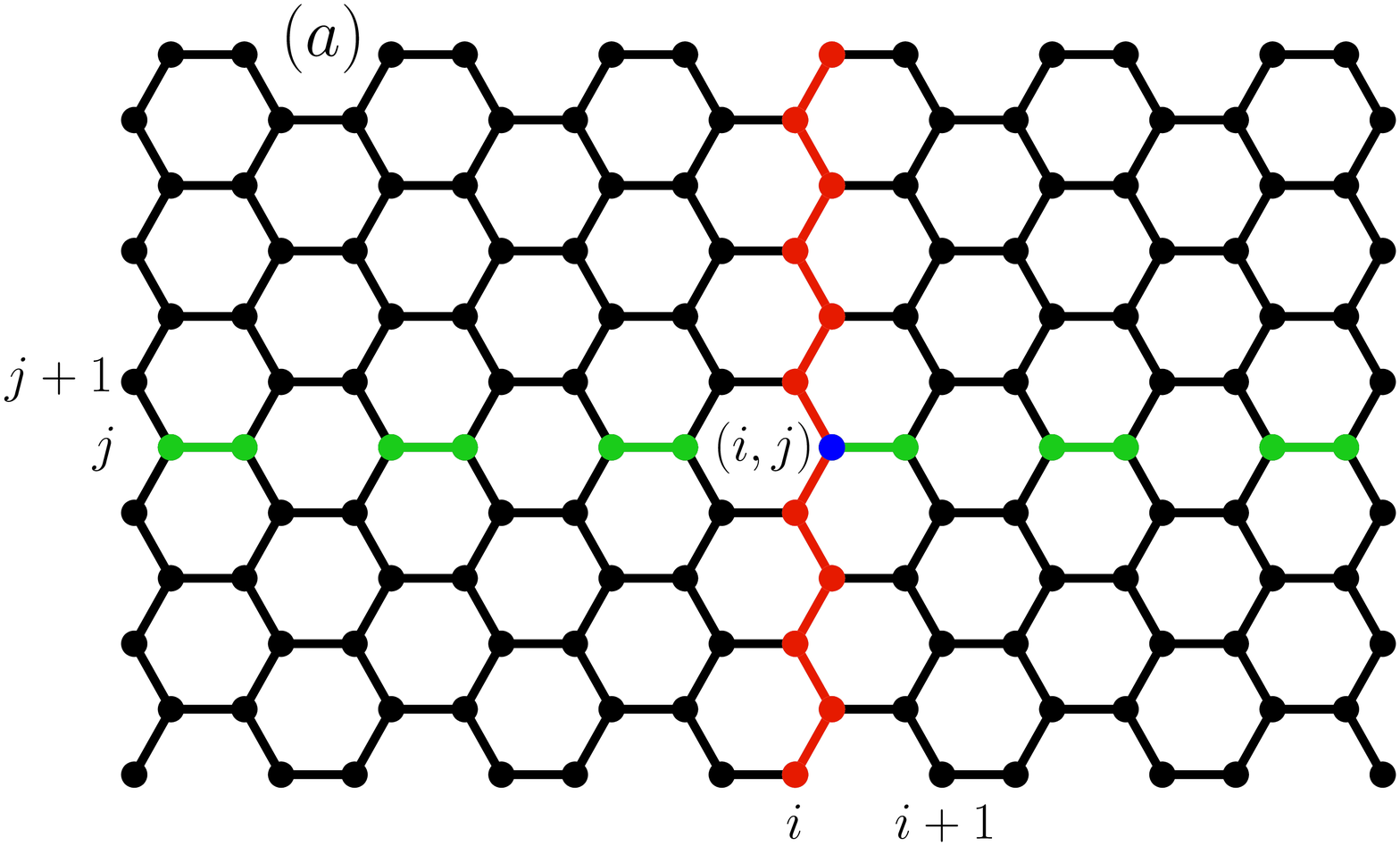}
    \includegraphics[width=0.22\textwidth,bb=0 0 240 270]{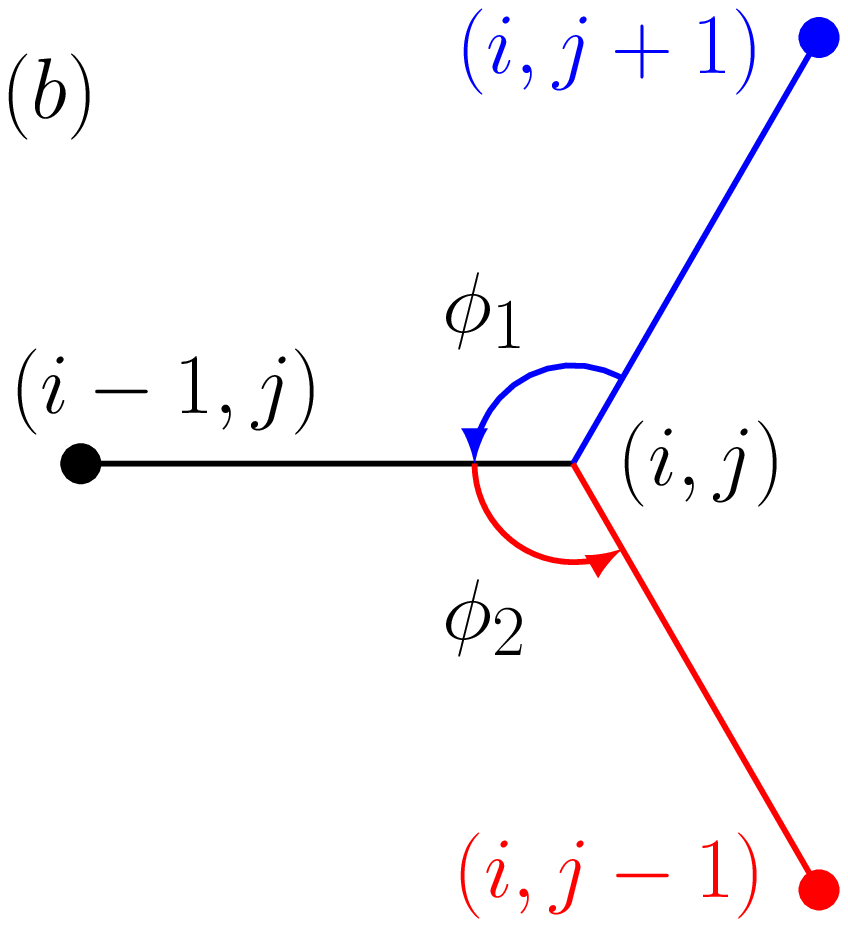}
    \hspace{0.6cm}
    \includegraphics[width=0.22\textwidth,bb=0 0 240 270]{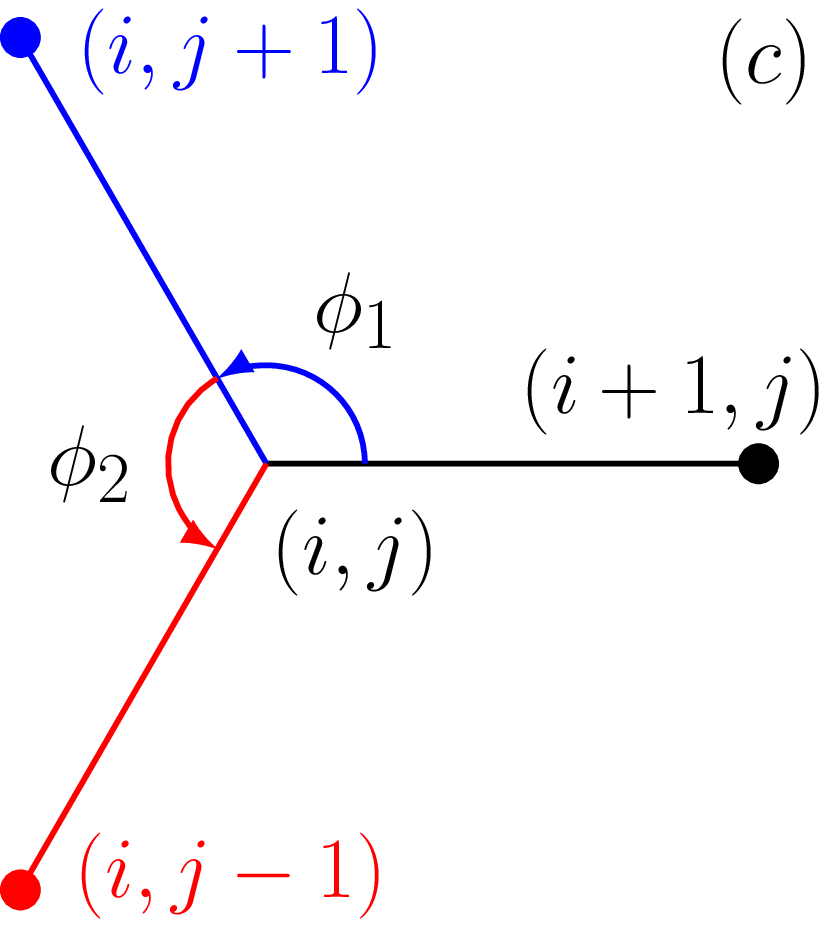}
    \caption{(a) Hexagonal structure of the graphene sheet. Each vertical zigzag chain is labelled by an index $i$, and the position along the chain labelled by an index $j$. Here for example the $i^{th}$ zigzag chain is highlighted in light red, and the $j^{th}$ atom in each chain is highlighted with light green. The atom at index $(i,j)$ is shown in dark blue. The two lower panels show the two possible cases with regard to the location of  neighboring atoms. Case $(b)$ corresponds to $i+j$ being an even integer (see text), and $\phi_1$ would be labelled as ${}_{i,j}\phi^{i-1,j}_{i,j+1}$. Case $(c)$ corresponds to $i+j$ being odd, and here $\phi_1$ would be labelled as ${}_{i,j}\phi^{i,j+1}_{i+1,j}$.}
    \label{fig:sheet_structure}
\end{figure}

The studied graphene structure at equilibrium is shown in Fig.~\ref{fig:sheet_structure}(a).
This arrangement gives rise to the graphene structure's characteristic honeycomb-like cell shape.
Each atom has three neighbors with the distance between two neighboring carbons $r_0 = 0.142~\text{nm}$ and the angle made by three consecutive carbons $\phi _0 = 2\pi/3 ~\text{rad}$ at equilibrium~\cite{KLGP13}.

Along the $i^{th}$ vertical zigzag chain shown in Fig.~\ref{fig:sheet_structure}(a), the $C_{i, j}$ carbon atom (where the
second index $j$ is numbering its position along the considered $i^{th}$ zigzag chain), shares one of its chemical bonds with either the $i-1$
zigzag chain to its left, when $i + j$ is an even integer [Fig.~\ref{fig:sheet_structure}(b)], or the $i+1$ zigzag chain in its right if $i + j$
is  odd [Fig.~\ref{fig:sheet_structure}(c)].
To calculate the position vector in the two dimensional plane $\boldsymbol{r}_{i, j}=\left(x_{i,j},y_{i,j}\right)^{T}$ with $^T$ denoting the transpose, of a carbon $C_{i, j}$ at equilibrium, which belongs to the $i^{th}$ zigzag chain and having the  $j^{th}$ position along the chain, we use
\begin{equation}
\boldsymbol{r}_{i, j} =
\begin{pmatrix}
(i - 1)r_0 \left[ \cos \left( \frac{\phi_0}{2}\right) +1 \right] +\frac{r_0}{2}\cos\left(\frac{\phi_0}{2}\right)\left[1-(-1)^{i+j}\right]\\
(j - 1)r_0 \sin \left(\frac{\phi _0}{2} \right)
\end{pmatrix}.
\label{eq:position_vec_r}
\end{equation}
Here, $r_{1, 1} = (0, 0)^T$ is considered as the origin of our cartesian coordinate system.
We define the vector from the carbon atom $C_{i, j}$ pointing to atom $C_{k, l}$ as $\boldsymbol{r}_{i, j}^{k, l} = (x_{k, l} - x_{i, j}, y_{k, l} - y_{i, j})^T$, and the angle between three neighboring atoms $C_{i, j}$, $C_{k, l}$ and $C_{m, n}$ is denoted as ${}_{i, j}{\phi _{k, l}^{m, n}}$. This angle is centered at point $(i, j)$ and is considered in an anticlockwise manner such that $\tan \left({}_{i, j}{\phi _{k, l}^{m, n}}\right) = \lVert \boldsymbol{r}_{i, j}^{k, l} \wedge \boldsymbol{r}_{i, j}^{m, n} \rVert / \boldsymbol{r}_{i, j}^{k, l} \cdot \boldsymbol{r}_{i, j}^{m, n} $,
where  $\boldsymbol{r}_{i, j}^{k, l} \wedge \boldsymbol{r}_{i, j}^{m, n}$ and $\boldsymbol{r}_{i, j}^{k, l} \cdot \boldsymbol{r}_{i, j}^{m, n}$ are respectively the usual wedge and dot product, with $\lVert \cdot \rVert$ being the usual Euclidian norm.
Through these definitions we obtain the graphene's geometrical description in a cartesian coordinate system where the calculations are straightforward.

\subsection{\label{subsec:governing_eq} Governing equations and computational aspects}

A displacement of an atom from its equilibrium position influences two potential energy terms in our two-dimensional force field.
The stretching of a covalent carbon-carbon bond is modelled via the Morse potential function~\cite{KLGP13}
\begin{equation}
V_s \left(r_{i, j}^{k, l}\right) = D \left[e^{-a (r_{i, j}^{k, l}- r_0) } -1\right]^2,
\label{eq:morse_potential}
\end{equation}
where $a = 1.96~\text{\AA}^{-1}$ is a constant such that its inverse is a characteristic length, $D = 5.7~\text{eV}$, i.e.~the depth of the potential, and $r_0$ is the equilibrium distance between two carbon atoms.
In addition, the bending of the angle created by three neighboring carbon atoms induces a potential energy~\cite{KLGP13}
 \begin{equation}
V_b\left(_{i, j}\phi _{k, l}^{m, n}\right) = \frac{d}{2}\left[ _{i, j}\phi _{k, l}^{m, n} - \phi _{0}\right]^2
- \frac{d^\prime}{3}\left[ _{i, j}\phi _{k, l}^{m, n} - \phi _{0} \right]^3,
\label{eq:bending_potential}
\end{equation}
where $d=7.0~\text{eV/rad}^2$ and $d^\prime = 4~\text{eV/rad}^3$  are the elastic and nonlinear parameters respectively.
In Eq.~\eqref{eq:bending_potential} $\phi _0$ is the equilibrium angle.
Consequently, we write the expression of the  Hamiltonian function (or energy of the system) as
\begin{align}
\nonumber
H = &\sum _{i, j} \frac{1}{2}m_{i, j} \left[ \left(\frac{dx_{i, j}}{dt}\right)^2 + \left(\frac{dy_{i, j}}{dt}\right)^2 \right] \\
\nonumber
&+ \sum _{\substack{i + j  \text{ even}}} \bigg\{ \frac{1}{2} \left[ V_s\left(r_{i, j}^{i, j+1}\right) + V_s\left(r_{i, j}^{i, j-1}\right) + V_s\left(r_{i, j}^{i-1, j}\right)\right] \\
\nonumber
&+  V_b \left( _{i, j}\phi _{i, j+1}^{i-1, j}\right) + V_b \left( _{i, j}\phi _{i-1, j}^{i, j - 1}\right) + V_b \left( _{i, j}\phi _{i, j - 1}^{i, j+1}\right) \bigg\}\\
\nonumber
&+ \sum _{\substack{i + j \text{ odd}}} \bigg\{ \frac{1}{2} \left[ V_s\left(r_{i, j}^{i, j+1}\right) + V_s\left(r_{i, j}^{i, j-1}\right) +V_s\left(r_{i, j}^{i + 1, j}\right)\right] \\
\nonumber
&+  V_b \left( _{i, j}\phi _{i + 1, j}^{i, j + 1}\right) + V_b \left( _{i, j}\phi _{i, j - 1}^{i + 1, j}\right) + V_b \left( _{i, j}\phi _{i, j + 1}^{i , j-1}\right) \bigg\}, \\
\label{eq:hamiltonian_graphene}
\end{align}
where $m_{i, j}$ is the mass of the carbon atom at site $(i, j)$.
For the carbon isotope ${}^{12}C$ the mass is taken to be $12.0~\text{amu}$ while for ${}^{13}C$ we are using $13.0~\text{amu}$.
Furthermore, in Hamiltonian~\eqref{eq:hamiltonian_graphene} the second summation accounts all the atoms where $i+j$ is even [Fig.~\ref{fig:sheet_structure}(b)] and the third summation accounts for those where $i+j$ is odd [Fig.~\ref{fig:sheet_structure}(c)].
Using the Hamiltonian formalism we find the equations of motion governing the evolution of $C_{i, j}$ in the cartesian coordinates system
\begin{equation}
m_{i, j} \frac{d^2 x_{i, j}}{dt^2} = - \frac{\partial H}{\partial x_{i, j}},\qquad m_{i, j} \frac{d^2 y_{i, j}}{dt^2} = - \frac{\partial H}{\partial y_{i, j}}.
\label{eq:eq_mot_xy_gen}
\end{equation}
The expressions in Eq.~\eqref{eq:eq_mot_xy_gen} are rather cumbersome since one has to take into consideration the orientation and the list of neighbors of a given carbon atom.
Note that, the equations of motion~\eqref{eq:eq_mot_xy_gen} conserve the total energy of the system $H$~\eqref{eq:hamiltonian_graphene}.

A small deviation from a trajectory in the phase space $\mathcal{S}$  has as coordinates the perturbations $\delta x_{i, j}$, $\delta y_{i, j}$, $\delta \dot{x}_{i, j}$ and $\delta \dot{y}_{i, j}$.
The displacements are measured in \AA\ and the velocities in \AA /ps.
This deviation vector evolves according to the so-called {\it variational equations}~\cite{S10,SG10}.
However, due to the highly complex equations of motion, the explicit writing of the variational equations is a very hard task.
For this reason, in order to compute the system's MLE we implement the so-called {\it two-particle method}~\cite{BGS76,MH17}, which consists of using the equations of motion~\eqref{eq:eq_mot_xy_gen}  to integrate an orbit with initial condition $\boldsymbol{X} (0) = (x_{i, j} (0), y_{i, j} (0), \dot{x}_{i, j}, (0) \dot{y}_{i, j} (0))$ along with a perturbed nearby orbit $\boldsymbol{X}^{\prime} (0)=  (x_{i, j} (0) + \delta x_{i, j} (0), y_{i, j} (0) +\delta y_{i, j} (0), \dot{x}_{i, j} (0) + \delta \dot{x}_{i, j} (0), \dot{y}_{i, j} (0) + \delta \dot{y}_{i, j} (0))$.
The deviation vector $\boldsymbol{v} (t)$ at any time $t$ of the evolution is thus obtained as $\boldsymbol{v} (t) = \boldsymbol{X}^{\prime} (t) - \boldsymbol{X} (t)$.
We then measure the averaged rate of exponential divergence of the two orbits $\boldsymbol{X}$ and $\boldsymbol{X}^{\prime}$ and compute the finite time maximum Lyapunov exponent~\cite{BGGS80a,BGGS80b,S10,PP16} (ftMLE)
\begin{equation}
\chi  = \frac{1}{t} \ln \left(\frac{||\boldsymbol{v}(t)||}{||\boldsymbol{v}(0)||} \right).
\label{eq:ftMLE}
\end{equation}
Then the system's MLE $\chi _1$ is given as $\chi _1  = lim _{t \rightarrow \infty} \chi $.
The MLE discriminates the system's orbits in a straightforward way: $\chi _1 > 0$ means that the  orbit is chaotic, while $\chi _1 = 0$ tells us that it is regular.
In addition, we note that the inverse of the MLE, referred as the  Lyapunov time $T_L=1/\chi_1$, provides a timescale of the system's chaotization, giving an estimate of how long the system takes to become chaotic \cite{S10}.

We solve the equations of motion using a symplectic integrator (SI).
SIs are extremely advantageous integration schemes designed especially for Hamiltonian systems as they exactly preserve the symplectic nature of the Hamiltonian dynamics.
One of their primary advantages is that they keep the error of the computed energy bounded and thus allow the utilization of a relatively large integration time step $\tau$, even when integrating for long times.
In our study we implement the  $ABA864$~\cite{BFLMM13} split SI of order four, which has proved to be very efficient~\cite{SS18,DMTS19} for two-dimensional classical systems.
The used integration time step $\tau = 0.06~\text{ps}$, keeps the relative energy error $|\left(H(t)-H(0)\right)/H(0)|$ below $10^{-7}$ and allows us to perform efficient and accurate computations of the ftMLE $\chi$ \eqref{eq:ftMLE} by using orbit perturbations with  $\lVert \boldsymbol{v} (0) \rVert \approx 10^{-6}$.
For lattices of a few thousand atoms (as in most of the cases considered here), the displacement components of $\boldsymbol{v}$ are thus on the order of $10^{-8}$\AA, and the velocity components similarly on the order of $10^{-8}$\AA/ps.
These are substantially smaller than the typical fluctuations in the positions and velocities of the atoms, which are on the order of $10^{-2}$\AA\ and $10^{-2}$\AA/ps respectively at the lowest energies considered here.
In our calculations of the ftMLE by the two-particle method, we renormalize the deviation vector to its initially considered norm value ($\lVert \boldsymbol{v} (0) \rVert \approx 10^{-6}$) after every picosecond in order to minimize numerical roundoff and overflow errors.
In addition, we carry out integrations up to a final time which is enough to ensure the convergence of the MLE estimator in all our simulations.
In most cases this time is  $t_{f}=10^{5}~\text{ps}$, but for the lowest considered energy density values, of $h=0.01$eV and $h=0.05$eV, for which the ftMLE shows a slower saturation, as well as in the cases shown in Fig.~\ref{fig:mle_linearised_system},  we use a final time of  $t_{f}=10^{6}~\text{ps}$.
Computations were run in parallel using OpenMP and GNU parallel \cite{T2018}.

It is worth mentioning that our calculations are performed on the microcanonical ensemble, where the system does not exchange energy with the exterior.
It is therefore more convenient to work with the mean energy density parameter $h = H/N$ which acts  as control parameter for our system, with $N$ being the total number of carbon atoms within the graphene shell.

\section{\label{sec:results} Numerical results}

\subsection{\label{subsec:graphene_sheet} Periodic Graphene}

In order to investigate the chaotic nature of the graphene shell, we consider a graphene sheet made of $N_i=60$ vertical zigzag chains, each possessing $N_j=48$ carbon atoms.
We further set periodic boundary conditions along the zigzag chains [vertical direction in Fig.~\ref{fig:sheet_structure}(a)] and the armchair edges [horizontal direction in Fig.~\ref{fig:sheet_structure}(a)] to mimic a bulk system.
As a good approximation of the natural composition of carbon materials, we primarily assume that the graphene shell is entirely made of ${}^{12}C$ isotopes.

\begin{figure}
    \centering
    \includegraphics[width=0.48\textwidth]{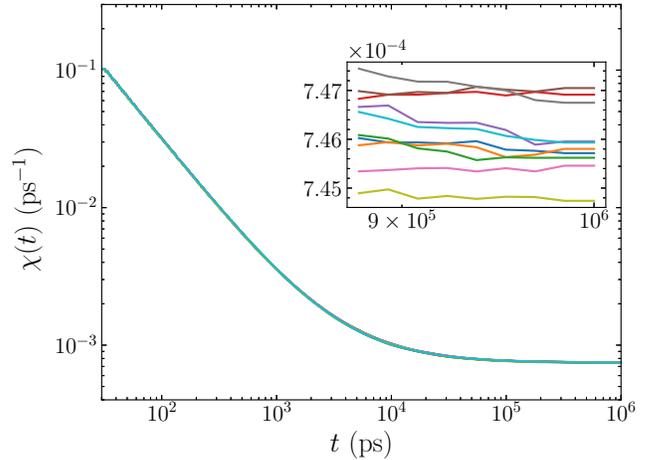}
    \caption{Time evolution of the ftMLE $\chi$ \eqref{eq:ftMLE} for 10 random initial conditions for the bulk graphene system of $N=N_i\cdot N_j=60 \cdot 48=2,880$ atoms with energy density $h=0.05$~eV in log-log scale. The different initial conditions produce very similar ftMLE values and all 10 curves practically overlap. Inset: A magnification of the final stage of the ftMLEs' evolution.}
    \label{fig:mle_sheet}
\end{figure}

Let us now explain how we estimate the system's MLE.
 In Fig.~\ref{fig:mle_sheet}  we see the convergence of the ftMLE $\chi$ \eqref{eq:ftMLE} obtained for  10 different sets of random initial conditions corresponding to a particular energy density, in this case $h = 0.05~\text{eV}$.
Each set corresponds to  randomly selected values of $\dot{x}_{i, j}(0)$ and $\dot{y}_{i, j} (0)$ which are compatible  with the particular energy density $h = 0.05~\text{eV}$. From the results of Fig.~\ref{fig:mle_sheet} we see that all  curves  practically overlap.
After an exponential decrease of the ftMLE at the earlier stages of the  evolution, a convergence to a value $\chi \approx 7.46\times 10^{-4}~\text{ps}^{-1}$ is observed, confirming the chaotic nature of interactions in the graphene shell. In the inset of Fig.~\ref{fig:mle_sheet} we see a magnification of the final stage of the ftMLEs' evolution, where the closeness of the results obtained from the different sets of initial conditions is evident.

We emphasize that the results of Fig.~\ref{fig:mle_sheet} are actually  independent of the type of the used initial conditions.
Different sets of initial set-ups, such as single or group site excitations on positions, momenta or both, produce practically the same results.
In addition, increasing the values of $N_i$ and $N_j$ does not affect the values of the ftMLE for the same energy density.

In our investigations we denote by $\bar{\chi}$  the average evolution of the ftMLE  \eqref{eq:ftMLE} over all the considered initial conditions, while $X_1$ represents the mean value of $\bar{\chi}$ over a time interval at the end of the integration, where the values of $\bar{\chi}$ have practically saturated at an almost constant value.
The uncertainties in evaluating both  $\bar{\chi}$ and $X_1$ are quantified by one standard deviation in the statistical process of their obtainment.

\begin{figure}
    \centering
    \includegraphics[width=0.48\textwidth]{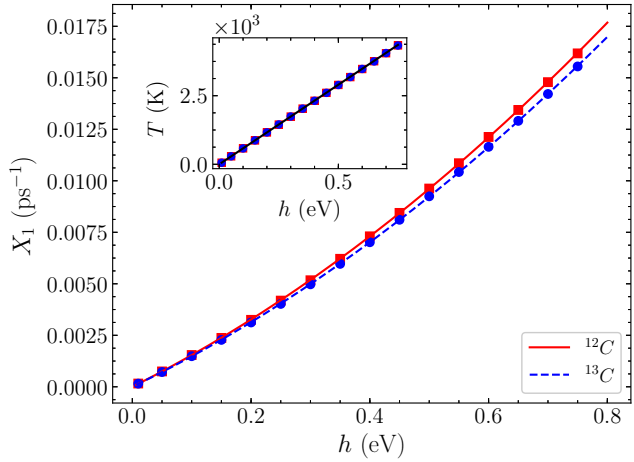}
    \caption{The average ftMLE $X_1$ of the graphene system of Fig.~\ref{fig:mle_sheet} as a function of the system's energy per particle $h$. The red squares correspond to results for graphene sheets composed solely of  ${}^{12}C$ isotopes, while the blue circles show results for ${}^{13}C$ atoms. The lines correspond to the analytical relation presented in Eq.~\eqref{eq:fitting_mle_sheet}. The very small error bars of the $X_1$ values (not visible at the plot's scales) means that the uncertainties of the measurements are very small. Inset: The temperature $T$ as a function of the energy density $h$ for both ${}^{12}C$ (red squares) and ${}^{13}C$ (blue circles). The straight line corresponds to the relation presented in Eq.~\eqref{eq:rel_en_dens_temp}.}
    \label{fig:mle_sheet_energy}
\end{figure}

Repeating the process of Fig.~\ref{fig:mle_sheet} for various values of the energy density $h$ we obtain the results of Fig.~\ref{fig:mle_sheet_energy} where the dependence of the MLE estimator $X_1$ on $h$ is depicted.
We see that for a graphene sheet composed solely of ${}^{12}C$ atoms (red squares) the MLE increases with increasing energy density  without showing any sign of a peculiar behavior, characteristic of potential structural instabilities, even at the largest values of $h$ considered here.
The error bars of the computed points do not appear in the graph because of their extremely small size.
At small values of the energy density, up to $h \approx 0.1$~eV, the MLE  is directly proportional to $h$.
For $h > 0.1$~eV, a quadratic correction adds to the linear initial behavior, giving rise to the parabola-like behavior observed at higher energy densities.
We have fitted~\cite{FITING19} the obtained results with the  quadratic function
\begin{equation}
X_1 (h) = \beta h + \gamma h^2,
\label{eq:fitting_mle_sheet}
\end{equation}
and obtained the values $\beta = 0.01447 \pm 0.00005~\text{ps}^{-1}\text{eV}^{-1}$ and $\gamma = 0.00951 \pm 0.00008~\text{ps}^{-1}\text{eV}^{-2}$ for the ${}^{12}C$ system (red curve). These coefficients do not practically change when
the system's size $N_i$ and/or $N_j$ increases. In addition, the results of MLE depicted in Fig.~\ref{fig:mle_sheet_energy} practically do not change if we include a $1.1\%$ doping of ${}^{13}C$ atoms into the graphene shell, which corresponds to its average concentration in naturally occurring carbon materials.
Indeed, even for the extreme case of a  lattice composed purely of ${}^{13}C$ isotopes (blue circles  in Fig.~\ref{fig:mle_sheet_energy}) the obtained parameters are $\beta = 0.01389 \pm 0.00004~\text{ps}^{-1}\text{eV}^{-1}$ and $\gamma = 0.00921 \pm 0.00007~\text{ps}^{-1}\text{eV}^{-2}$, which are relatively close to their values observed for the ${}^{12}C$ lattice. The slightly smaller chaoticity of the  ${}^{13}C$ shell can likely be attributed to the fact that  ${}^{13}C$ atoms have higher masses and therefore greater inertia and consequent stability in the lattice.

In our simulations the system's Lyapunov time $T_L$, i.e.~the inverse of the MLE, is on the order of $10^2$ to $10^4$ps.
Since the characteristic frequencies of the optical phonon modes of the graphene model~\cite{PCCP} are on the order of $10^{14}$Hz, the corresponding vibrational time scales are on the order of $10^{-2}$ps.
Comparing this with $T_L$, we  see that in graphene it takes more than $10^4$ oscillations of the high frequency modes before chaos sets in.
Thus chaotization is a relatively slow process in graphene.

In the inset of Fig.~\ref{fig:mle_sheet_energy} we see the validity of the linear relation
\begin{equation}
h = 2k_B T,
\label{eq:rel_en_dens_temp}
\end{equation}
between the energy density $h$ and the temperature $T$ (in Kelvin - K), with $k_B= 8.617 \times 10^{-5}~\text{eV K}^{-1}$ being the Boltzmann constant, of the considered two-dimensional system even for the largest $T$ or $h$ values investigated. The red squares and the blue circles correspond to data obtained from our simulations for, respectively, the ${}^{12}C$ and ${}^{13}C$ graphene sheet, while the straight line represents Eq.~\eqref{eq:rel_en_dens_temp}. We note that in our computations we estimate the temperature $T$ as the mean value of the quantity $H_c/(k_BN)$, where $H_c$ is the system's kinetic energy and $N=N_i\times N_j$ is the total number of carbon atoms. This quantity is computed after the kinetic energy of the system reaches equilibrium, and it is averaged over time as well over ten different initial conditions.

\begin{figure}[tb]
    \centering
    \includegraphics[width=0.48\textwidth]{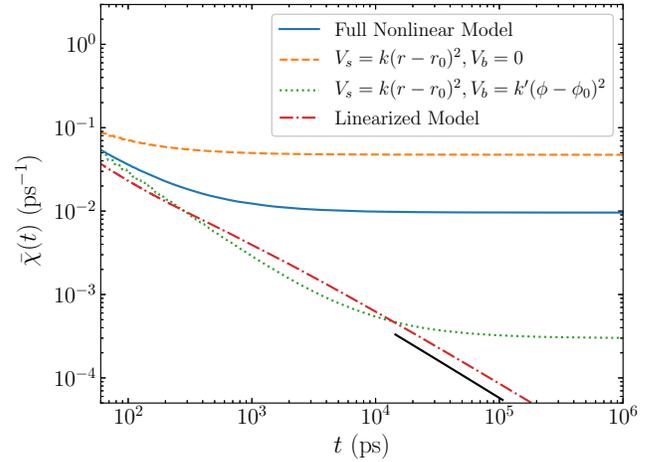}
    \caption{A comparison of the evolution of the average ftMLE $\bar{\chi}$ at $h=0.5$eV considering various modifications of the used model. Results are shown for the complete nonlinear potential of Eqs.~\eqref{eq:morse_potential} and \eqref{eq:bending_potential} (blue solid curve), only harmonic stretching and bending potentials (green dotted curve), solely stretching harmonic coupling (orange dashed curve), and the pure linearized version of the total potential (red dash-dotted curve). Each case is averaged over 10 realizations. The black curve indicates a function proportional to  $\ln(t)/t$ (See text).}
    \label{fig:mle_linearised_system}
\end{figure}

It is also of interest that the chaoticity in graphene does not arise solely as a result of the potential's nonlinearity, but the 2D geometry itself is sufficient to produce chaotic behavior. This is demonstrated in Fig.~\ref{fig:mle_linearised_system}, where a comparison of the ftMLE computation for different modifications of the full nonlinear system at a relatively large energy density ($h=0.5$~eV) is shown. We see that chaotic  behavior persists even when approximating the Morse potential (Eq.~\ref{eq:morse_potential}) as a harmonic coupling and only taking the quadratic term from the bending potential in Eq.~\eqref{eq:bending_potential}, i.e.~setting $d^\prime=0$. The ftMLE of this system tends to a positive value (green dotted curve in Fig.~\ref{fig:mle_linearised_system}), which nevertheless is almost two orders of magnitude smaller than the ftMLE of the full nonlinear system (blue solid curve). In fact, due to the 2D geometry of the system, even if we only consider a harmonic stretching interaction between neighboring atoms (omitting any angular terms)  we observe a significantly strong chaotic behavior (orange dashed curve). The comparison of the orange and green curves in Fig.~\ref{fig:mle_linearised_system} clearly reveals the stabilizing effect of the bending potential, as the addition of the quadratic angular potential significantly decreases the value of the MLE. It is worth noting that the genuine linearization of the system, through first-order approximations of the forces near the equilibrium state (red dash-dotted curve), results to a decrease of the ftMLE towards 0 at a rate of $\ln(t)/t$ (denoted by the black solid curve in Fig.~\ref{fig:mle_linearised_system})  as expected for regular motions (see e.g.~Sect.~5.3 of Ref.~\cite{S10} and references therein).

\subsection{\label{subsec:nanoribbon} Graphene Nanoribbons}

GNRs are finite width strips of graphene with free edges on the two opposite sides defining the width.
We model an {\it armchair GNR} by considering periodic boundary conditions on the zigzag edges [horizontal direction in Fig.~\ref{fig:sheet_structure}(a)], while for the {\it zigzag GNR} we apply periodic boundary conditions into the armchair edges [vertical direction in Fig.~\ref{fig:sheet_structure}(a)].
In both cases, free boundary conditions are considered for the non-periodic edges.

\begin{figure}
    \centering
    \includegraphics[width=0.48\textwidth]{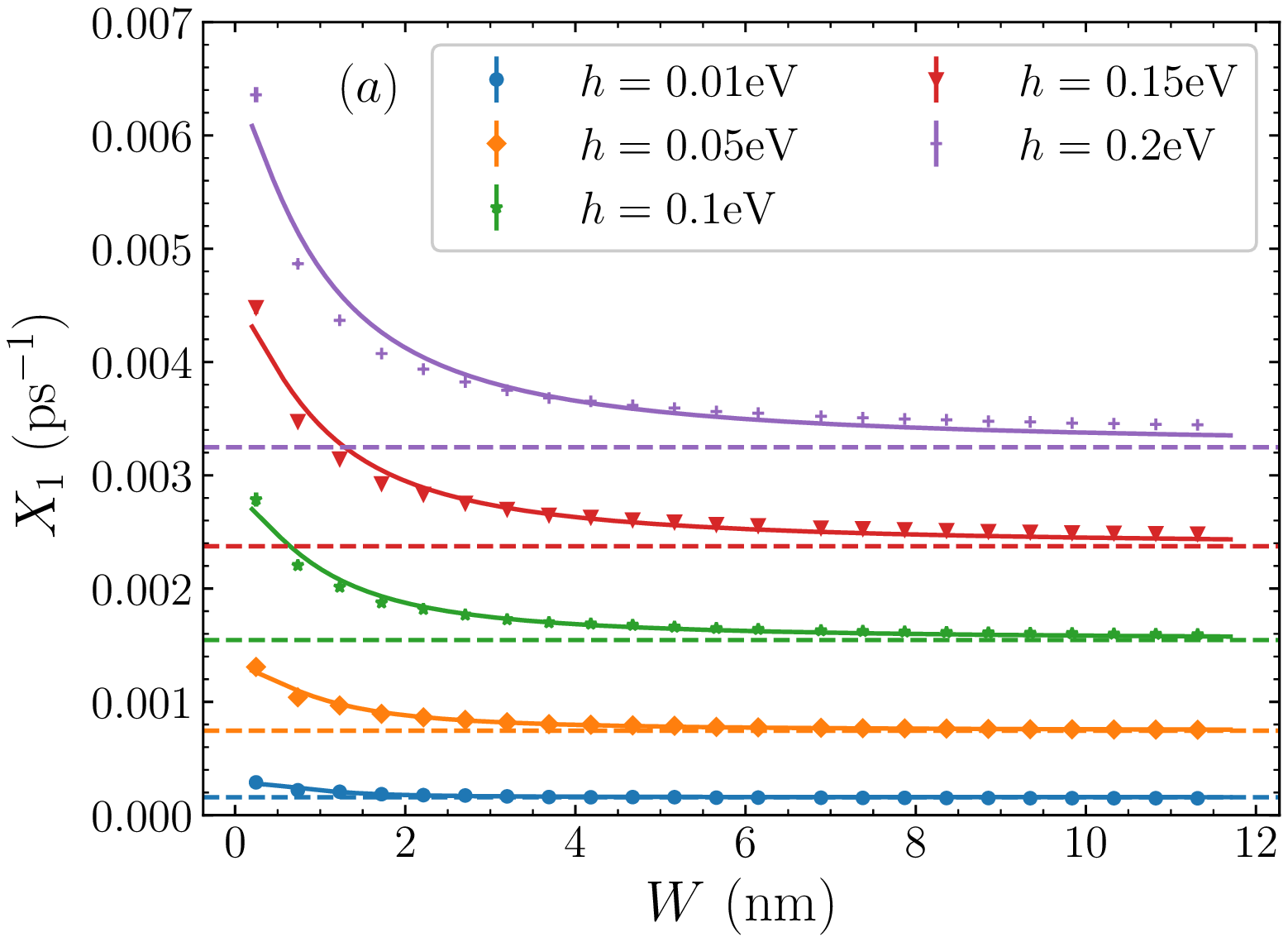}
    \includegraphics[width=0.48\textwidth]{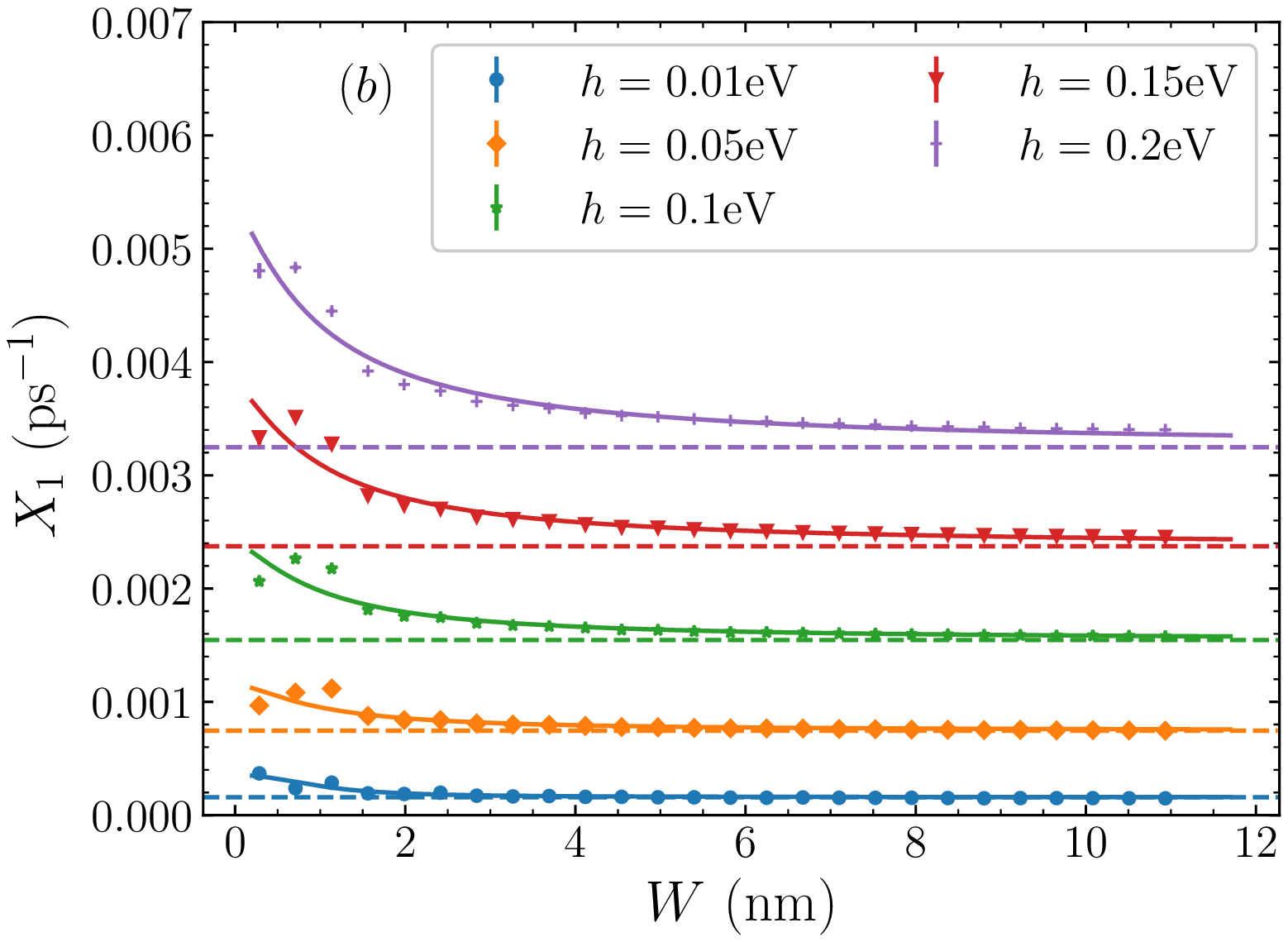}
    \caption{The MLE estimator $X_1$ of (a) armchair and (b) zigzag nanoribbons as a function of GNR's width $W$, for different energy densities $h$. The horizontal dashed lines asymptotically below each data set represent the $X_1$ value of a periodic graphene sheet, with the same energy per particle, depicted in Fig.~\ref{fig:mle_sheet_energy}. The solid curves correspond to the fitting of the data with the function of Eq.~\ref{eq:fitting_mle_ribbon}. Note that once again, the error bars on the data points are extremely small and not visible on the scale of the plot.}
    \label{fig:mle_ribbon}
\end{figure}

Here we investigate the influence of the GNR's width $W$ on the chaoticity of the ribbon structure.
The width of an armchair GNR with $N_j$ carbon atoms is $W_{A}={\sqrt{3}}(N_j-1)r_0/{2}$, while in the case of a zigzag GNR with $N_i$ atoms the width is $W_Z=({3N_i}/{2}-1)r_0$.
Fig.~\ref{fig:mle_ribbon} shows the variation of the MLE estimator $X_1$ with $W$, at various energy densities $h$ for the armchair [Fig.~\ref{fig:mle_ribbon}(a)] and zigzag [Fig.~\ref{fig:mle_ribbon}(b)] GNRs.
For both nanoribbon types the $X_1$ values increase with increasing energy densities, and larger values are observed compared to the periodic graphene systems (see horizontal dashed lines in Fig.~\ref{fig:mle_ribbon}).
Overall, armchair GNRs exhibit a slightly more chaotic behavior than zigzag GNRs.

For armchair GNRs a decrease of $X_1$ with $W$ always appears, meaning that wider GNRs are less chaotic and consequently more stable. The same behavior is generally observed for zigzag GNRs as well, apart from perhaps the 1-2 smallest
values of $W$ where at some energy densities a non monotonic behavior can be found [Fig.~\ref{fig:mle_ribbon}(b)].

In order to quantify our findings, the data of Fig.~\ref{fig:mle_ribbon} are fitted with a decreasing Hill function with an added
constant term of the form
\begin{equation}
    X_1 \left(W\right) = \frac{A}{1 + W^n} + X_1^b,
    \label{eq:fitting_mle_ribbon}
\end{equation}
where the parameters $A$ and $n$ are free to be fitted, the ribbon width $W$ is expressed in nm in this formula,
and $X_1^b$ is the estimated value of the MLE of the bulk graphene sheet with periodic boundary conditions corresponding to the particular energy density, as shown in Fig.~\ref{fig:mle_sheet_energy}.
Thus $X_1^b$ represents a limiting value characterizing the asymptotic behavior of Eq.~\eqref{eq:fitting_mle_ribbon},
as one expects that at very large values of $W$ the GNR's MLE should approach that of bulk graphene.
This expected behavior is evidently observed in the data of Fig.~\ref{fig:mle_ribbon}.

\begin{figure}
    \centering
    \includegraphics[width=0.48\textwidth]{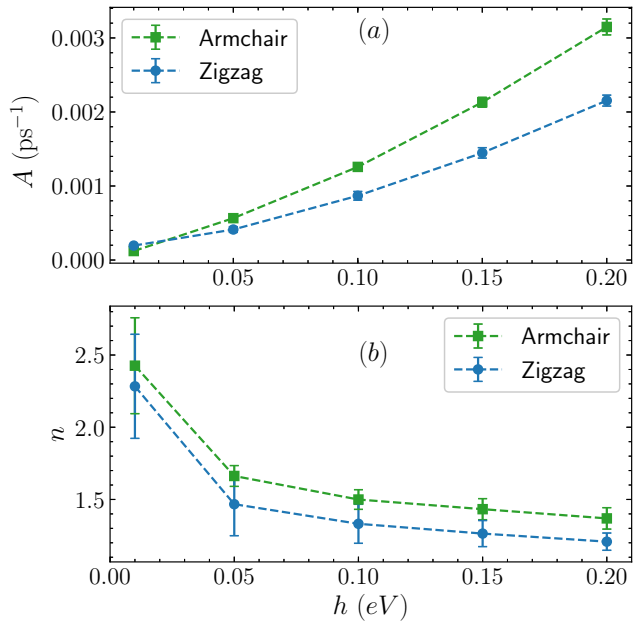}
    \caption{Variation of the  fitting parameters (a) $A$ and (b) $n$ of Eq.~\eqref{eq:fitting_mle_ribbon}, with the energy per particle $h$, for armchair GNRs (green squares) and zigzag ribbons (blue circles). The lines connect the points to guide the eye.}
    \label{fig:fit_params}
\end{figure}

Fig.~\ref{fig:fit_params} shows the variation of the fitting parameters $A$ and $n$ of Eq.~\eqref{eq:fitting_mle_ribbon} with energy density $h$.
We see that the overall behavior for the two GNR types is similar.
The coefficient $A$ increases with energy density, while it is generally larger in the armchair case.
This is in agreement with the larger range of $X_1$ values observed for armchair GNRs in Fig.~\ref{fig:mle_ribbon}.
The exponent $n$ decreases with $h$, reflecting a less steep relative decrease of $X_1$ with $W$ for larger energy densities.
Armchair GNRs correspond to larger values of $n$ than zigzag GNRs of the same energy density,
describing a relatively more abrupt decrease in this case. Regardless of the value of $n$, there is eventually a saturation
of the MLE estimator $X_1$ given in Eq.~(\ref{eq:fitting_mle_ribbon}) at the $X_1^b$ value obtained for the bulk graphene sheet.

\section{\label{sec:conclusion}Summary and concluding remarks}

Using a two-dimensional Hamiltonian model for describing the dynamics of planar graphene structures we investigated the material's chaoticity through the computation of the system's MLE. In our study we have considered perfect graphene crystals, as well as both zigzag and armchair graphene nanoribbons.

We found that in all cases the MLE increases with energy density and we showed that the MLE of graphene sheets does not practically change in the presence of different carbon isotopes ${}^{12}C$ or ${}^{13}C$ in the structure.
Chaos in graphene is revealed after more than $10^4$ oscillations of the characteristic normal modes.
Furthermore, due the 2D geometry of the system, even harmonic interaction potentials between atoms is enough to produce chaos.

Our findings show that the edge effects from the free boundaries in the GNRs result in a more chaotic behavior than is observed in the bulk structure. The MLE values of GNRs decrease as their width increases, tending asymptotically to the values observed in the case of the perfect graphene crystal. Furthermore, we have found armchair GNRs to be slightly more chaotic than the zigzag ribbons.

We expect that the investigation of the chaotic behavior of graphene structures performed in this work would be extended in the future by considering the impact of various in- or out-of-plane defects, where in the latter case torsional energy terms and out of plane motions of carbon atoms should be taken into account~\cite{PCCP,EPJB}.

\begin{acknowledgments}
M.~H., B.~M.~M. and Ch.~S.~acknowledge support by the National Research Foundation (NRF) of South Africa. M.~H.~was also supported by the Deutsche Akademische Austauschdienst (DAAD). G.~K.~and Ch.~S.~were supported by the Erasmus+/International Credit Mobility KA107 program.
We thank the High Performance Computing facility of the University of Cape Town~\cite{ICTS19}, the Center for High Performance Computing~\cite{CHPC19} of South Africa, and the Center for Information Services and High Performance Computing (ZIH) at the Technische Universit\"at Dresden for providing computational resources for this project.
We also thank A. Ngapasare and M. Werner for useful discussions and the two anonymous referees for their comments, which helped us improve the presentation of our work.
\end{acknowledgments}

\section*{Data Availability}
The data that support the findings of this study are available from the corresponding author upon reasonable request.

\nocite{*}

\begin{thebibliography}{999}

    \bibitem{NGMJZDGF04} K.~S.~Novoselov, A.~K.~Geim, S.~V.~Morozov, D.~Jiang, Y.~Zhang, S.~V.~Dubonos, I.~V.~Grigorieva  and A.~A.~Firsov, Science {\bf 306}, 666 (2004).

\bibitem{RMP09}  A. H. Castro Neto, F. Guinea, N. M. R. Peres, K.~S.~Novoselov, and A.~K.~Geim, Rev. Mod. Phys. {\bf 81}, 109 (2009).

	\bibitem{ZMCLSPR10} Y.~Zhu, S.~Murali, W.~Cai, X.~Li, J.~W.~Suk, J.~R.~Potts, and R.~S.~Ruoff, Adv. Mater. {\bf 22}, 3906 (2010).
	
    \bibitem{SJZDKS11} V.~Singh, D.~Joung, L.~Zhai, S.~Das, S.~I.~Khondaker and S.~Seal, Prog. Mater. Sc. {\bf 56}, 1178 (2011).

    \bibitem{NGMJKGDF05} K.~S.~Novoselov, A.~K.~Geim, S.~Morozov, D.~Jiang, M.~I.~Katsnelson, I.~Grigorieva, S.~Dubonos and  A.~A.~Firsov, Nature {\bf 438}, 197 (2005).

    \bibitem{GN07} A.~K.~Geim and K.~S.~Novoselov, Nat. Mater. {\bf 6} 183 (2007).

    \bibitem{GCTPNBBM08} S.~Ghosh, I.~Calizo, D.~Teweldebrhan, E.~P.~Pikatilov, D.~L.~Nika, A.~A.~Balandin, W.~Bao, F.~Miao and C.~N.~Lau, Appl.~Phys.~Lett. {\bf 92}, 151911 (2008).

	\bibitem{BGBCTML08}	A.~A.~Balandin, S.~Ghosh, W.~Bao, I.~Calizo, D.~Teweldebrhan, F.~Miao and C.~N.~Lau, Nano Lett. {\bf 8}, 902 (2008).

    \bibitem{LWKH08} C. Lee, X. Wei, J. W. Kysar, and J. Hone, Science {\bf 321}, 385 (2008).
	
	\bibitem{LYC12} J.~-U.~Lee, D.~Yoon, and H.~Cheong, Nano Lett. {\bf 12}, 4444 (2012).

    \bibitem{TPPJFGNG09} G.~Tsoukleri, J.~Parthenios, K.~Papagelis, R.~Jalil, A.~C.~Ferrari, A.~K.~Geim, K.~S.~Novoselov and C.~Galiotis, Small {\bf 5}, 2397 (2009).

 \bibitem{papagACSNano10}  O.~Frank, G.~Tsoukleri, J.~Parthenios, , K.~Papagelis, I. Riaz, R. Jalil, K.~S.~Novoselov and C.~Galiotis, ACS Nano {\bf 4}, 3131 (2010).

 \bibitem{revJPCM15} C.~Daniels, A.~Horning, A.~Phillips, D.~V.~P.~Massote, L.~Liang, Z.~Bullard, B.~G.~Sumpter and V.~Meunier,
 J.~Phys.: Cond.~Matt. {\bf 27}, 373002 (2015).





	\bibitem{BKB12} D.~A.~Brownson, D.~K.~Kampouris and C.~E.~Banks, Chem. Soc. Rev {\bf 41}, 6944 (2012).
	
	\bibitem{sens} F.~Schedin, A.~K.~Geim, S.~V.~Morozov, E.~W.~Hill, P.~Blake, M.~I.~Katsnelson and K.~S.~Novoselov, Nat.
Mater. {\bf 6}, 652 (2007).
	
	\bibitem{YYZYC11} H.~J.~Yoon, J.~H.~Yang, Z.~Zhou, S.~S.~Yang and M.~M.~C.~Cheng,  Sens. Actuator B-Chem. {\bf 157}, 310 (2011).


 \bibitem{reson} J.~S.~Bunch, A.~M.~Van Der Zande, S.~S.~Verbridge, I.~W.~Frank, D.~M.~Tanenbaum, J.~M.~Parpia,
H.~G.~Craighead and P.~L.~McEuen, Science {\bf 315}, 490 (2007).


\bibitem{avouris} Y.-M.~Lin, A.~Valdes-Garcia, S.-J.~Han, D.~B.~Farmer, I.~Meric, Y.~Sun, Y.~Wu, C.~Dimitrakopoulos, A.~Grill,
P.~Avouris and K.~A.~Jenkins, Science {\bf 332}, 1294 (2011).

	\bibitem{MCKLPCA12} C.~Mattevi, F.~Colléaux, H.~Kim, Y.~H.~Lin, K.~T.~Park, M.~Chhowalla and T.~D.~Anthopoulos, Nanotechnology {\bf 23}, 344017 (2012).
	
	\bibitem{LYNZJ10} C.~Liu, Z.~Yu, D.~Neff, A.~Zhamu and B.~Z.~Jang Nano Lett. {\bf 10}, 4863 (2010).

	\bibitem{ZXZLZ10} L.~Zhang, J.~Xia, Q.~Zhao, L.~Liu and Z.~Zhang, Small {\bf 6}, 537 (2010).
	
    \bibitem{ZM10} Z.~Xu and M.~J.~Buehler, ACS Nano {\bf 4}, 3869 (2010).

    \bibitem{nscrolls} S.~Zhu and T.~Li, J.~Phys.~D: Appl.~Phys.~{\bf 46}, 075301 (2013).

    \bibitem{SSPG14} A.~Sgouros, M.~M.~Sigalas, K.~Papagelis and G.~Kalosakas, J.~Phys.: Condens.~Matter {\bf 26}, 125301 (2014).

    \bibitem{ncages} L.~Zhang, X.~Zeng and X.~Wang, Sci.~Rep.~{\bf 3}, 3162 (2013).

    \bibitem{SKSP15} A.~P.~Sgouros, G.~Kalosakas, M.~M.~Sigalas and K.~Papagelis, RSC Adv. {\bf 5}, 39930 (2015).

    \bibitem{GZG09} Z.~Guo, D.~Zhang and X.-G.~Gong, Appl. Phys. Lett. {\bf 95}, 163103 (2009).

    \bibitem{HRC09} J.~Hu, X.~Ruan and Y.~P.~Chen, Nano Lett. {\bf 9}, 2730 (2009).

    \bibitem{RuoffNatMat12}  S.~Chen, Q.~Wu, C.~Mishra, J.~Kang, H.~Zhang, K.~Cho, W.~Cai, A.~A.~Balandin and R.~S.~Ruoff,
    Nat.~Mat.~{\bf 11}, 203 (2012).

    \bibitem{CK12} L.~Chen and S.~Kumar, J.~Appl.~Phys. {\bf 112}, 043502 (2012).

    \bibitem{ZD12} Z.~G.~Fthenakis and D.~Tomanek, Phys. Rev. B {\bf 86}, 125418 (2012).

    \bibitem{KSLCBT13} N.~Khosravian, M.~K.~Samani, G.~C.~Loh, G.~C.~K.~Chen, D.~Baillargeat and B.~K.~Tay, Comp. Mat. Sc. {\bf 79}, 132 (2013).

    \bibitem{LCYZ13} X.~Li, J.~Chen, C.~Yu and G.~Zhang, Appl.~Phys.~Lett. {\bf 103}, 013111 (2013).

    \bibitem{YLZYW13} P.~Yang, X.~Li, Y.~Zhao, H.~Yang and S.~Wang, Phys.~Lett.~A {\bf 377}, 2141 (2013).

    \bibitem{YZ13} C.~Yu and G.~Zhang, J.~Appl.~Phys. {\bf 113}, 044306 (2013).

    \bibitem{ZD14} Z.~G.~Fthenakis, Z. Zhu and D.~Tomanek, Phys. Rev. B {\bf 89}, 125421 (2014).

    \bibitem{ZHY15} J.~Zhang, Y.~Hong and Y.~Yue, J.~Appl.~Phys {\bf 117}, 134307 (2015).

    \bibitem{SKH10} A.~V.~Savin, Y.~S.~Kivshar and B.~Hu, Phys.~Rev.~B {\bf 82}, 195422 (2010).

    \bibitem{LT12} G.~C.~Loh, E.~H.~T~Teo and B.~K.~Tay, Diamond Relat. Mater. {\bf 23}, 88 (2012).

    \bibitem{IMLB13} M.~Z.~Islam, M.~Mahboob, R.~L.~Lowe and S.~.E.~Bechtel, J.~Phys.~D: Appl. Phys. {\bf 46}, 435302 (2013).

    \bibitem{KKGP15} E.~N.~Koukaras, G.~Kalosakas, C.~Galiotis and K.~Papagelis, Sci.~Rep. {\bf 5}, 12923 (2015).

    \bibitem{kopidakis}  G.~Kopidakis, C.~Z.~Wang, C.~M.~Soukoulis and K.~M.~Ho, J.~Phys.: Cond.~Matt.~{\bf 9}, 7071 (1997).

    \bibitem{AKPKPG15} Ch.~Androulidakis, E.~N.~Koukaras, J.~Parthnios, G.~Kalosakas, K.~Papagelis and C.~Galiotis, Sci.~Rep. {\bf 5}, 18219 (2015).

    \bibitem{phlifetime}  Z.~Wei, J.~Yang, K.~Bi and Y.~Chen, J.~Appl.~Phys.~{\bf 116}, 153503 (2014).

     \bibitem{ZMA09} H.~Zhao, K.~Min, and N.~R.~Aluru, Nano Lett. {\bf 9}, 3012 (2009).

    \bibitem{TT10} J.~L.~Tsai, J.-F.~Yu, Mater. Design {\bf 31}, 194 (2010).

    \bibitem{CQ13} A.~Cao and J.~Qu, Appl.~Phys.~Lett. {\bf 102}, 071902 (2013).

    \bibitem{KLGP13} G.~Kalosakas, N.~N.~Lathiotakis, C.~Galiotis and K.~Papagelis, J. Appl. Phys. {\bf 113}, 134307 (2013).

    \bibitem{ZMFZPLLGZZAZL14} P.~Zhang, L.~Ma, F.~Fan, Z.~Zeng, C.~Peng, P.~E.~Loya, Z.~Liu, Y.~Gong, J.~Zhang, X.~Zhang, P.~M.~Ajayan, T.~Zhu and J.~Lou, Nat. Comm. {\bf 5}, 3782 (2014).

    \bibitem{sgouros2dmat} A.~P.~Sgouros, G.~Kalosakas, C.~Galiotis and K.~Papagelis, 2D Mater. {\bf 3}, 025033 (2016).

    \bibitem{NP10} M.~Neek-Amal and F.~M.~Peeters, Appl.~Phys.~Lett. {\bf 97}, 153118 (2010).

    \bibitem{SKPG18} A.~P.~Sgouros, G.~Kalosakas, K.~Papagelis and C.~Galiotis, Sci. Rep. {\bf 8}, 9593 (2018).

    \bibitem{SGL16} Ch.~Skokos, G.~A.~Gottwald and J.~Laskar, Chaos Detection and Predictability, Lecture Notes in Physics {\bf 915}, Springer Verlag  (2016).

    \bibitem{S10} Ch.~Skokos, Lect.~Notes~Phys. {\bf 790}, 63 (2010).

    \bibitem{PP16} A.~Pikovsky and A.~Politi, Lyapunov exponents: a tool to explore complex dynamics, Cambridge University Press (2016).

    \bibitem{BD01}  J.~Barre and T.~Dauxois, Europhys. Lett. {\bf 55}, 164 (2001).
	
	\bibitem{HKSS19} M.~Hillebrand, G.~Kalosakas, A.~Schwellnus and Ch.~Skokos, Phys. Rev. E {\bf 99}, 022213 (2019).
	
    \bibitem{PCCP}  Z.~G.~Fthenakis, G.~Kalosakas, G.~D.~Chatzidakis, C.~Galiotis, K.~Papagelis and N.~N.~Lathiotakis,
    Phys.~Chem.~Chem.~Phys.~{\bf 19}, 30925 (2017).

    \bibitem{EPJB}  G.~D.~Chatzidakis, G.~Kalosakas, Z.~G.~Fthenakis and N.~N.~Lathiotakis, Eur.~Phys.~J.~B {\bf 91}, 11 (2018).
	
    \bibitem{SG10} Ch.~Skokos and E.~Gerlach,  Phys.~Rev.~E, {\bf 82}, 036704 (2010).

    \bibitem{BGS76} G.~Bennetin, L.~Galgani and J.-M. Strelcyn, Phys.~Rev.~A {\bf 14}, 2338 (1976).

	\bibitem{MH17} L.~Mei and L.~Huang, Comp.~Phys.~Comm. {\bf 224}, 108 (2018).

	\bibitem{BGGS80a} G.~Benettin, L.~Galgani, A.~Giorgilli and J.-M.~Strelcyn, Meccanica {\bf 15}, 9 (1980).

    \bibitem{BGGS80b} G.~Benettin, L.~Galgani, A.~Giorgilli and  J.-M.~Strelcyn, Meccanica {\bf 15}, 21 (1980).

    \bibitem{BFLMM13} S.~Blanes, F.~Casas, A.~Farres, J.~Laskar, J.~Makazaga and A.~Murua, Appl.~Numer.~Math. {\bf 68}, 58 (2013).

	\bibitem{SS18} B.~Senyange, Ch.~Skokos, Eur.~Phys.~J.~Spec.~Top., ~{\bf 227}, 625 (2018).
	
	\bibitem{DMTS19} C.~Danieli, B.~Many Manda, T.~Mithun and Ch.~Skokos, Math.~in Eng.~{\bf 1}, 447 (2019).

    \bibitem{T2018} O.~Tange, GNU Parallel 2018, Ole Tange, \url{https://doi.org/10.5281/zenodo.1146014}

    \bibitem{FITING19} The fit was performed using the Levenberg-Marquardt algorithm for least squares optimisation \cite{M77}.

    \bibitem{M77} J.~J.~More, The Levenberg-Marquardt Algorithm: Implementation and Theory, Lecture Notes in Mathematics {\bf 630}, Springer Verlag (1977).
	
    \bibitem{ICTS19} \url{http://hpc.uct.ac.za}.

    \bibitem{CHPC19}  \url{https://www.chpc.ac.za}.








	
















\end{thebibliography}

\end{document}